# On the Possible Trajectories of Spinning Particles.
# I. Free Particles

## A. N. Tarakanov [*]

By means of the method of moving Frenet-Serret frame the set of equations of motion is derived for spinning particle in an arbitrary external field, which is determined by potential depending from both position and the state of movement, as well as by two pseudo-vectors one of which is easily associated with external magnetic field, and another still remains undetermined. The equations give a possibility to describe the motion of both massive and massless particles with spin. All solutions of the equations of motion in the absence of external fields were found, and besides, we give more precise definition of a free object. It turns out that the massive particles always possess a longitudinal polarization. There are possible transversal motions of the following types: 1) oscillatory motion with proper frequency, 2) circular motion, and 3) complicated motion along rosette trajectories round the center of inertia with the velocity, varying in the limits $v_{\min} < v < v_{\max}$. Free massless particles can either fluctuate or move along complicated paths around fixed centers of balance, when the spin of the particles can have any direction.

## 1. Introduction

Explanation of electrodynamic phenomena by means of an interaction of moving electrical charges goes back to the theory of Weber which has analyzed experiments of Ampère and has established «*the fundamental law of electrical action*» for the force of interaction of moving charges, depending both on the relative position of charges, and on their state of a motion [1], [2]. Alternatives of the Weber's theory, distracting from how much they have appeared adequate, were considered by Gauss ([3], S. 616), Riemann ([4], S. 326), Clausius [5]-[7], Edlund [8], while Tisserand [9] and Zöllner [10] have extended the Weber's law to gravitation, believing that laws of Newton and Coulomb are only approximations of the more general Weber's law which should be applied to interaction of atoms which structure was not known yet about. If these theories are to be considered as first steps in an establishing of real electromagnetic interaction, then Maxwell has initially refused from any "assumptions as to the physical nature of electricity, or adding anything to that which has been already proved by experiment", using hydrodynamic analogy for electromagnetic field ([11], p. 30). In equations of motion with force which Lorentz has added to the Maxwell's equations, charged particles are assumed to be pointwise so their proper angular momentum is not considered. After the discovery of electron spin there was a necessity of its account in equations of motion that has partially been made within framework of special relativity and quantum electrodynamics (see, e. g., [12], [13]). It became clear after theoretical discovery of Zitterbewegung by Schrödinger [14] and numerous papers devoted to it, that the spin makes essential impact on a trajectory even for the free particles. However a possibility of taking account of spin in terms of classical nonrelativistic theory has not been properly analyzed.

In the papers [15]-[18] an idea was proposed for the description of objects with internal degrees of freedom (spinning particles) by generalization of the Second Newton's Law, and classic non-relativistic theory of such objects interacting with external fields has been developed. Pseudo-vectors $\mathbf{S} = \varsigma\mathbf{s} + \mathbf{S}^{\text{ext}}$ and $\mathbf{C} = -\varsigma\Omega_0^2\mathbf{s} + \mathbf{C}^{\text{ext}}$, included in the equation of motion, contain pseudo-vector of spin $\mathbf{s}$, representing internal degrees of freedom of the object, and pseudo-vectors $\mathbf{S}^{\text{ext}}$ and $\mathbf{C}^{\text{ext}}$ determined by external fields. Constant $\varsigma$ has dimension of the inverse square of the velocity dimension. When $\varsigma = -c^{-2}$, where $c$ is the velocity of light, equation of motion reduces to non-relativistic limit of the Frenkel-Matisson-Weyssenhoff equation [19]-[22].

---

[*] Electronic address: tarak-ph@mail.ru

Consequence of the equation of motion is equation of energy balance, which is an integral of motion only at certain condition.

In [17]-[18] solutions are found for non-relativistic equation of motion for free mass point in the center-of-inertia reference frame (r. f.), which generally does not coincide with the center-of-mass r. f. moving through the complicated trajectory around the direction of movement of the center of inertia that can be interpreted as Zitterbewegung. In this connection a particle should be considered as non-inertial r. f. with some structure, whose spin is classical proper angular momentum. The main conclusion, which was done in [15]-[18], lies in the fact that the electric charge should not be seen as a physical quantity characterizing electromagnetic interaction, but as a consequence of the presence of the particle spin. Then electromagnetic interaction can be interpreted as interaction of the spin with external field. So the question arises, how the motion of spinning particles in an electromagnetic field and their interaction with each other can be associated with the behavior of charged particles considered in the framework of the standard (classical or quantum) electrodynamics.

The structure of this paper is as follows. In Sec. 2 the equations of motion in any external field are obtained. It follows that pseudo-vector $\mathbf{C}^{\text{ext}}$ represents an external magnetic field. Using the method of moving Frenet-Serret frame equations of motion reduce to the set of equations, applied for obtaining of trajectories of both massive and massless particles. It is shown in Sec. 3 that the definition of "free object" allows existence of non-zero potential function. All solutions of free equation of motion determining by conditions $\partial U / \partial \mathbf{R} = \mathbf{0}$, $\mathbf{S}^{\text{ext}} = \mathbf{0}$, $\mathbf{C}^{\text{ext}} = \mathbf{0}$ are obtained. Interestingly, the equation of motion admits solutions corresponding fluctuations of massless particles with arbitrary spin direction about the fixed center of balance. Conclusive Sec. 4 includes short discussion of the results. Appendix contains information on the moving Frenet-Serret frame.

## 2. Equations of Motion in Arbitrary External Field

The equation of motion of a particle with spin in an external field can be written as the Second Newton's Law $d\mathbf{P} / dt = \mathbf{F}$ ([15]-[18]), or

$$\frac{d}{dt}\left\{m_0 \mathbf{V} + \varsigma[\mathbf{s} \times \dot{\mathbf{V}}]\right\} + \varsigma\Omega_0^2[\mathbf{s} \times \mathbf{V}] = -\frac{\partial U}{\partial \mathbf{R}} + \frac{d}{dt}\left(\frac{\partial U}{\partial \mathbf{V}}\right) + [\mathbf{C}^{\text{ext}} \times \mathbf{V}] - \frac{d}{dt}[\mathbf{S}^{\text{ext}} \times \dot{\mathbf{V}}], \quad (2.1)$$

where

$$\mathbf{F} = \mathbf{F}_g + \mathbf{F}^{\text{ext}}, \quad (2.2)$$

is a force, containing the gyroscopic force,

$$\mathbf{F}_g = -\varsigma\Omega_0^2[\mathbf{s} \times \mathbf{V}], \quad (2.3)$$

arising due to non-inertial object, and external force

$$\mathbf{F}^{\text{ext}} = -\frac{\partial U}{\partial \mathbf{R}} + [\mathbf{C}^{\text{ext}} \times \mathbf{V}]. \quad (2.4)$$

Here

$$U = U_0 + \varsigma\Omega_0^2([\mathbf{R} \times \mathbf{V}] \cdot \mathbf{s}) - ([\mathbf{R} \times \mathbf{V}] \cdot \mathbf{C}^{\text{ext}}) \quad (2.5)$$

is potential energy of interaction, $\mathbf{\Omega}_0$ is Zitterbewegung frequency of free particle.

Dynamic momentum of the object

$$\mathbf{P} = \mathbf{P}_{\text{kin}} + \mathbf{A} = m_0\mathbf{V} + \varsigma[\mathbf{s} \times \dot{\mathbf{V}}] - \frac{\partial U}{\partial \mathbf{V}} + [\mathbf{S}^{\text{ext}} \times \dot{\mathbf{V}}], \quad (2.6)$$

contains proper kinetic momentum

$$\mathbf{P}_{\text{kin}} = m_0\mathbf{V} + \varsigma[\mathbf{s} \times \dot{\mathbf{V}}], \quad (2.7)$$

which does not depend on external fields (that's why we should include here the term $\varsigma[\mathbf{s} \times \dot{\mathbf{V}}]$), and addition

$$\mathbf{A} = -\frac{\partial U}{\partial \mathbf{V}} + [\mathbf{S}^{ext} \times \dot{\mathbf{V}}], \tag{2.8}$$

arising due to interaction with external fields (that's why we should not include in $\mathbf{A}$ the term $\varsigma[\mathbf{s} \times \dot{\mathbf{V}}]$). This addition leads to a renormalization of the force $\mathbf{F}$, so that the change of the proper kinetic momentum (2.7) is determined by the right-hand side of the equation of motion (2.1), which can be compared with the Lorentz force, if we put

$$\mathbf{E} = -\frac{\partial U}{\partial \mathbf{R}} + \frac{d}{dt}\left(\frac{\partial U}{\partial \mathbf{V}} - [\mathbf{S}^{ext} \times \dot{\mathbf{V}}]\right) = -\frac{\partial U}{\partial \mathbf{R}} - \frac{d\mathbf{A}}{dt}, \tag{2.9}$$

$$\mathbf{B} = -\mathbf{C}^{ext}, \tag{2.10}$$

which shows that addition (2.8) has a role of vector potential, introduced by F. E. Neumann ([23], [24]). However, as distinct from standard definition of the electric field (2.9) contains total derivative of the vector $\mathbf{A}$ with respect to time, but not the partial one.

Assuming magnetic field has to be caused exclusively by moving charges, i. e. by particles with spin, for determination of magnetic induction $\mathbf{B}$ in terms of interaction potential one should consider the problem of two interacting particles with spin, what will be done in a separate paper. For the purpose of this article it is enough to accept the identification of $\mathbf{C}^{ext}$ with magnetic induction, according to (2.10).

If we assume the definition (2.9)-(2.10), the equation of motion (2.1) takes the form

$$\frac{d}{dt}(m_0\mathbf{V} + \varsigma[\mathbf{s} \times \dot{\mathbf{V}}]) = \mathbf{E} + [\mathbf{V} \times (\mathbf{B} + \varsigma\Omega_0^2\mathbf{s})], \tag{2.11}$$

coinciding with non-relativistic equation for a charged particle in electric and magnetic field at $\mathbf{s} = \mathbf{0}$.

The equation of motion of the spin is given by

$$\dot{\mathbf{s}} = [\mathbf{\Omega} \times \mathbf{s}], \tag{2.12}$$

where $\mathbf{\Omega}$ is angular velocity of spin precession.

Equation (2.1) leads to the equation of energy balance

$$\frac{d\mathcal{E}}{dt} = \frac{\partial U}{\partial t} + \sum_{k=0}^{N}\left(\frac{\partial U}{\partial \dot{\mathbf{V}}^{(k)}} \cdot \dot{\mathbf{V}}^{(k+1)}\right), \tag{2.13}$$

where

$$\mathcal{E} = \frac{m_0\mathbf{V}^2}{2} + \varsigma(\mathbf{V} \cdot [\mathbf{s} \times \dot{\mathbf{V}}]) + (\mathbf{V} \cdot [\mathbf{S}^{ext} \times \dot{\mathbf{V}}]) - (\mathbf{V} \cdot \frac{\partial U}{\partial \mathbf{V}}) + U \tag{2.14}$$

is total energy of the object.

It is necessary to say a few words about the structure of potential function (2.5), defining an interaction of the object with fields produced by external sources. Every source can be considered as a collection of point sources, characterized by the radius vectors $\mathbf{R}_k$ and their derivatives $\mathbf{V}_k = \dot{\mathbf{R}}_k, \dot{\mathbf{V}}_k, \ldots$, $k = 1, 2, \ldots, K$, where $K$ is the number of sources. Since the potential function is a scalar, its dependence on time, coordinates, velocity and accelerations can be expressed as

$$U = U(t; |\mathbf{R} - \mathbf{R}_1|, |\mathbf{R} - \mathbf{R}_2|, \ldots |\mathbf{R} - \mathbf{R}_K|; |\mathbf{V} - \mathbf{V}_1|, \ldots, |\mathbf{V}^{(N)} - \mathbf{V}_K^{(N)}|) =$$
$$= \sum_{k=1}^{K} U_k(t; |\mathbf{R} - \mathbf{R}_k|, |\mathbf{V} - \mathbf{V}_k|, \ldots, |\mathbf{V}^{(N)} - \mathbf{V}_k^{(N)}|), \tag{2.15}$$

where $U_k$ is a potential function of interaction of $k$-th point source with the object in question. Representation of the potential function as the sum of (2.15) may be considered as an expression of the principle of superposition: the contribution of each point source in the interaction does not depend on contributions from other sources.

In view of (2.15) we have

$$\frac{\partial U}{\partial \mathbf{R}} = \sum_{k=1}^{K} \frac{\partial U_k}{\partial |\mathbf{R}-\mathbf{R}_k|} \frac{\mathbf{R}-\mathbf{R}_k}{|\mathbf{R}-\mathbf{R}_k|}, \quad \frac{\partial U}{\partial \mathbf{V}} = \sum_{k=1}^{K} \frac{\partial U_k}{\partial |\mathbf{V}-\mathbf{V}_k|} \frac{\mathbf{V}-\mathbf{V}_k}{|\mathbf{V}-\mathbf{V}_k|}, \dots \quad (2.16)$$

If all point sources are fixed, then $\mathbf{V}_k = \mathbf{0}$ and

$$\frac{\partial U}{\partial \mathbf{V}} = \sum_{k=1}^{K} \frac{\partial U_k}{V \partial V} \mathbf{V} = \frac{\partial U}{V \partial V} \mathbf{V}. \quad (2.17)$$

Let a r. f. $K'$, whose origin in K is specified by radius vector $\mathbf{R}_{(K')}$, be the r. f. moving relative to absolute one with the velocity $\mathbf{V}_{(K')}$. Let's represent the equation (2.1), written down in absolute r. f. K, in moving r. f. $K'$. Then

$$\mathbf{R} = \mathbf{R}_{(K')} + \mathbf{r}, \quad \mathbf{V} = \mathbf{V}_{(K')} + \mathbf{v}, \dots, \quad (2.18)$$

where $\mathbf{r}$ and $\mathbf{v}$ are radius vector and velocity of the object relative to the origin of $K'$. It is easy to see that relations (2.16) in $K'$ are the same form as in K, i. e.

$$\frac{\partial U}{\partial \mathbf{R}} = \sum_{k=1}^{K} \frac{\partial U_k}{\partial |\mathbf{r}-\mathbf{r}_k|} \frac{(\mathbf{r}-\mathbf{r}_k)}{|\mathbf{r}-\mathbf{r}_k|} = \frac{\partial U}{\partial \mathbf{r}}, \quad \frac{\partial U}{\partial \mathbf{V}} = \sum_{k=1}^{K} \frac{\partial U_k}{\partial |\mathbf{v}-\mathbf{v}_k|} \frac{(\mathbf{v}-\mathbf{v}_k)}{|\mathbf{v}-\mathbf{v}_k|} = \frac{\partial U}{\partial \mathbf{v}}, \dots \quad (2.19)$$

Substitution of (2.18), (2.19) into (2.1) leads to the form

$$\frac{d}{dt}(m_0 \mathbf{v} + \varsigma[\mathbf{s} \times \dot{\mathbf{v}}]) + \varsigma \Omega_0^2 [\mathbf{s} \times \mathbf{v}] + \frac{d}{dt}(m_0 \mathbf{V}_{(K')} + \varsigma[\mathbf{s} \times \dot{\mathbf{V}}_{(K')}]) + \varsigma \Omega_0^2 [\mathbf{s} \times \mathbf{V}_{(K')}] =$$
$$= -\frac{\partial U}{\partial \mathbf{r}} + \frac{d}{dt}\frac{\partial U}{\partial \mathbf{v}} + [\mathbf{v} \times \mathbf{B}] - \frac{d}{dt}[\mathbf{S}^{\text{ext}} \times \dot{\mathbf{v}}] + [\mathbf{V}_{(K')} \times \mathbf{B}] - \frac{d}{dt}[\mathbf{S}^{\text{ext}} \times \dot{\mathbf{V}}_{(K')}]. \quad (2.20)$$

There is no necessity to use vector potential to describe the motion in a constant electromagnetic field. In this case electric field is defined as

$$\mathbf{E} = -\frac{\partial U}{\partial \mathbf{R}} = -\frac{\partial U}{\partial \mathbf{r}}, \quad (2.21)$$

whence it follows

$$U = -\int (\mathbf{E} \cdot d\mathbf{R}) + u(\mathbf{V}, \dot{\mathbf{V}}, \ddot{\mathbf{V}}, \dots, \dot{\mathbf{V}}^{(N)}) = -\int (\mathbf{E} \cdot d\mathbf{r}) + u(\mathbf{v}, \dot{\mathbf{v}}, \ddot{\mathbf{v}}, \dots, \dot{\mathbf{v}}^{(N)}), \quad (2.22)$$

where $u(\mathbf{v}, \dot{\mathbf{v}}, \ddot{\mathbf{v}}, \dots, \dot{\mathbf{v}}^{(N)}) = 0$ for the rest particle.

On the other hand, in the case of varying field its dependence on time may be considered in potential function (2.15) together with conservation of the definition (2.21) instead of generally accepted definition $\mathbf{E} = -\nabla U - \partial \mathbf{A}/c\partial t$ in Maxwell electrodynamics. Then the equation (2.1) will become

$$\frac{d}{dt}\left(m_0 \mathbf{V} - \frac{\partial u}{\partial \mathbf{V}} + \varsigma[\mathbf{s} \times \dot{\mathbf{V}}] + [\mathbf{S}^{\text{ext}} \times \dot{\mathbf{V}}]\right) = \mathbf{E} + [\mathbf{V} \times (\mathbf{B} + \varsigma \Omega_0^2 \mathbf{s})], \quad (2.23)$$

which coincides with classical equation if conditions $\mathbf{s} = \mathbf{0}$, $\partial u / \partial \mathbf{V} - [\mathbf{S}^{\text{ext}} \times \dot{\mathbf{V}}] = \mathbf{const}$ are fulfilled. As a result (2.20) turns into

$$\frac{d}{dt}\left(m_0 \mathbf{v} - \frac{\partial u}{\partial \mathbf{v}} + \varsigma[\mathbf{s} \times \dot{\mathbf{v}}] + [\mathbf{S}^{\text{ext}} \times \dot{\mathbf{v}}]\right) + \varsigma \Omega_0^2 [\mathbf{s} \times \mathbf{v}] +$$
$$+ \frac{d}{dt}\left(m_0 \mathbf{V}_{(K')} + \varsigma[\mathbf{s} \times \dot{\mathbf{V}}_{(K')}] + [\mathbf{S}^{\text{ext}} \times \dot{\mathbf{V}}_{(K')}]\right) + \varsigma \Omega_0^2 [\mathbf{s} \times \mathbf{V}_{(K')}] = \mathbf{E} + [\mathbf{v} \times \mathbf{B}] + [\mathbf{V}_{(K')} \times \mathbf{B}]. \quad (2.24)$$

In the moving r. f. $K'$ relations $\mathbf{V}_{(K')} = \mathbf{0}$, $\dot{\mathbf{V}}_{(K')} = \mathbf{0}$ should be fulfilled. Therefore equation (2.24) is splitted into two equations

$$\frac{d}{dt}\left(m_0 \mathbf{v} - \frac{\partial u}{\partial \mathbf{v}} + \varsigma[\mathbf{s} \times \dot{\mathbf{v}}] + [\mathbf{S}^{\text{ext}} \times \dot{\mathbf{v}}]\right) + \varsigma \Omega_0^2 [\mathbf{s} \times \mathbf{v}] = \mathbf{E} + [\mathbf{v} \times \mathbf{B}], \quad (2.25)$$

$$\frac{d}{dt}\left(m_0\mathbf{V}_{(K')} + \varsigma[\mathbf{s}\times\dot{\mathbf{V}}_{(K')}] + [\mathbf{S}^{\text{ext}}\times\dot{\mathbf{V}}_{(K')}]\right) + \varsigma\Omega_0^2[\mathbf{s}\times\mathbf{V}_{(K')}] = [\mathbf{V}_{(K')}\times\mathbf{B}]. \quad (2.26)$$

We shall introduce in $K'$ orthonormal basis (A.1)-(A.3), where the velocity $\mathbf{V}$ should be replaced by $\mathbf{v}$. Then any vector or pseudo-vector can be expanded in this basis. Hence

$$\mathbf{\Omega} = \Omega_\tau\mathbf{e}_\tau + \Omega_n\mathbf{e}_n + \Omega_b\mathbf{e}_b, \quad (2.27)$$

$$\mathbf{s} = s_\tau\mathbf{e}_\tau + s_n\mathbf{e}_n + s_b\mathbf{e}_b = \left(s_\tau - s_n\frac{(\mathbf{v}\cdot\dot{\mathbf{v}})}{|[\mathbf{v}\times\dot{\mathbf{v}}]|}\right)\frac{\mathbf{v}}{v} + s_n\frac{v\dot{\mathbf{v}}}{|[\mathbf{v}\times\dot{\mathbf{v}}]|} + s_b\frac{[\mathbf{v}\times\dot{\mathbf{v}}]}{|[\mathbf{v}\times\dot{\mathbf{v}}]|}, \quad (2.28)$$

and it follows from (2.12) that $\mathbf{s}^2 = s^2 = s_\tau^2 + s_n^2 + s_b^2 = \text{const}$. Substitution of (2.28) into (2.12) gives equations

$$\dot{s}_\tau = \Omega_n s_b + (vK - \Omega_b)s_n, \quad (2.29)$$

$$\dot{s}_n = (\Omega_b - vK)s_\tau + (vT - \Omega_\tau)s_b, \quad (2.30)$$

$$\dot{s}_b = (\Omega_\tau - vT)s_n - \Omega_n s_\tau, \quad (2.31)$$

only two of which are independent since any equation is derived from a linear combination of the other two. If (2.27) is the Darboux vector (A.9), $\mathbf{\Omega} = \mathbf{\Omega}_D$, then (2.29)-(2.31) imply that spin components $s_\tau$, $s_n$ and $s_b$ are constant. In addition, (2.28) leads to the relation $(\mathbf{v}\cdot[\mathbf{s}\times\dot{\mathbf{v}}]) = -s_b|[\mathbf{v}\times\dot{\mathbf{v}}]|$.

We choose the r. f. $K'$ so that its velocity was orthogonal to the plane of vectors $\mathbf{v}$ and $\dot{\mathbf{v}}$. When it is inertial frame, then $\dot{\mathbf{V}}_{(K')} = \mathbf{0}$. In this case the binormal direction is kept constant

$$\mathbf{V}_{(K')} = V_{(K')}\frac{[\mathbf{v}\times\dot{\mathbf{v}}]}{|[\mathbf{v}\times\dot{\mathbf{v}}]|} = V_{(K')}\mathbf{e}_b, \quad (2.32)$$

i. e., as it follows from (A.8), the torsion vanishes

$$T = \frac{(\mathbf{v}\cdot[\dot{\mathbf{v}}\times\ddot{\mathbf{v}}])}{[\mathbf{v}\times\dot{\mathbf{v}}]^2} = 0, \quad (2.33)$$

which greatly simplifies the equation of motion.

For the sake of simplicity we assume that the source of the field $U$ is at rest and function $u$ does not depend on accelerations $\dot{\mathbf{v}}, \ddot{\mathbf{v}}, ..., \dot{\mathbf{v}}^{(N)}$ of the particle. Then

$$\frac{d}{dt}\frac{\partial u}{\partial \mathbf{v}} = \frac{d}{dt}\left(\frac{du}{vdv}\mathbf{v}\right) = \frac{du}{vdv}\dot{\mathbf{v}} + \frac{d}{dt}\left(\frac{du}{vdv}\right)\mathbf{v} = \frac{du}{vdv}\dot{\mathbf{v}} + (\mathbf{v}\cdot\dot{\mathbf{v}})\frac{d}{dv}\left(\frac{du}{vdv}\right)\mathbf{v}. \quad (2.34)$$

We seek a solution of equation (2.25) in the form

$$\mathbf{v}(t) = v(t)\mathbf{e}_\tau = v(t)[\cos\Phi(t)\mathbf{e}_X + \sin\Phi(t)\mathbf{e}_Y], \quad (2.35)$$

where $\mathbf{e}_X$, $\mathbf{e}_Y$ are unit vectors of coordinate system in the plane which is orthogonal to the Z axis in absolute r. f., $\mathbf{e}_Z = \mathbf{e}_b$. In view of the spin components $s_\tau$, $s_n$, $s_b$ can be either positive or negative, we can assume that $\dot{\Phi} \geq 0$. Then the spin of the particle according to (2.28) is equal to

$$\mathbf{s} = (s_\tau\cos\Phi - s_n\sin\Phi)\mathbf{e}_X + (s_\tau\sin\Phi + s_n\cos\Phi)\mathbf{e}_Y + s_b\mathbf{e}_Z. \quad (2.36)$$

In view of (2.34), (2.36), (A.23) equations of motion (2.25), (2.26) and spin precession (2.12) are

$$\frac{d}{dt}\left[\left(m_0 v - \frac{du}{dv} - (\varsigma s_b + S_Z^{\text{ext}})v\dot\Phi\right)\cos\Phi - (\varsigma s_b + S_Z^{\text{ext}})\dot v\sin\Phi\right]\mathbf{e}_X +$$

$$+\frac{d}{dt}\left[\left(m_0 v - \frac{du}{dv} - (\varsigma s_b + S_Z^{\text{ext}})v\dot\Phi\right)\sin\Phi + (\varsigma s_b + S_Z^{\text{ext}})\dot v\cos\Phi\right]\mathbf{e}_Y -$$

$$-\varsigma s_b v\Omega_0^2[\sin\Phi\mathbf{e}_X - \cos\Phi\mathbf{e}_Y] + \left[\varsigma s_\tau(v\ddot\Phi + \dot v\dot\Phi) - \varsigma s_n(\ddot v + \Omega_0^2 v)\right]\mathbf{e}_Z + \quad (2.37)$$

$$+\frac{d}{dt}\left[(S_X^{\text{ext}}\dot v + S_Y^{\text{ext}}v\dot\Phi)\sin\Phi + (S_X^{\text{ext}}v\dot\Phi - S_Y^{\text{ext}}\dot v)\cos\Phi\right]\mathbf{e}_Z =$$

$$= (E_X + vB_Z\sin\Phi)\mathbf{e}_X + (E_Y - vB_Z\cos\Phi)\mathbf{e}_Y + (E_Z + vB_Y\cos\Phi - vB_X\sin\Phi)\mathbf{e}_Z,$$

$$\frac{d}{dt}\left[\dot V_{(K')}(S_Y^{\text{ext}} + \varsigma s_\tau\sin\Phi + \varsigma s_n\cos\Phi)\right]\mathbf{e}_X + \left[\varsigma s_\tau\sin\Phi + \varsigma s_n\cos\Phi\right]\Omega_0^2 V_{(K')}\mathbf{e}_X -$$

$$-\frac{d}{dt}\left[\dot V_{(K')}(S_X^{\text{ext}} + \varsigma s_\tau\cos\Phi - \varsigma s_n\sin\Phi)\right]\mathbf{e}_Y + \left[\varsigma s_n\sin\Phi - \varsigma s_\tau\cos\Phi\right]\Omega_0^2 V_{(K')}\mathbf{e}_Y + \quad (2.38)$$

$$+m_0\dot V_{(K')}\mathbf{e}_Z = -V_{(K')}B_Y\mathbf{e}_X + V_{(K')}B_X\mathbf{e}_Y,$$

$$\dot{\mathbf{s}} = [\mathbf{\Omega}_D \times \mathbf{s}] = \dot\Phi\left[-(s_\tau\sin\Phi + s_n\cos\Phi)\mathbf{e}_X + (s_\tau\cos\Phi - s_n\sin\Phi)\mathbf{e}_Y\right], \quad (2.39)$$

where

$$\mathbf{\Omega}_D = \Omega_b\mathbf{e}_b = vK\mathbf{e}_b = \frac{[\mathbf{v}\times\dot{\mathbf{v}}]}{v^2} = \dot\Phi\mathbf{e}_Z. \quad (2.40)$$

It follows from (2.38), that $\dot V_{(K')} = 0$ at $m_0 \neq 0$, i. e. moving r. f. $K'$ is inertial one, $V_{(K')} = \text{const}$, so that (2.38) is equivalent to equations

$$\varsigma s_n\Omega_0^2 = B_X\sin\Phi - B_Y\cos\Phi, \quad (2.41)$$

$$\varsigma s_\tau\Omega_0^2 = -B_X\cos\Phi - B_Y\sin\Phi, \quad (2.42)$$

whence it follows

$$\varsigma^2(s_\tau^2 + s_n^2)\Omega_0^4 = B_X^2 + B_Y^2, \quad (2.43)$$

$$\sin\Phi = \frac{\varsigma\Omega_0^2(s_n B_X - s_\tau B_Y)}{B_X^2 + B_Y^2}, \quad \cos\Phi = -\frac{\varsigma\Omega_0^2(s_\tau B_X + s_n B_Y)}{B_X^2 + B_Y^2}. \quad (2.44)$$

Note that the relation (2.43) contradicts to the condition $\Omega_0 = \text{const}$ when the projection of the magnetic field $\mathbf{B}$ onto the plane (XY) is not constant. Therefore, in general, it should be assumed that either the frequency $\Omega_0$ may be variable, and its dependence, as well as the same dependence of function $\Phi$, on time and coordinates are determined by the relations (2.43) and (2.44), respectively, or $s_\tau = s_n = 0$, $B_X = B_Y = 0$.

In the case $m_0 = 0$ the moving r. f. $K'$ may be non-inertial one. Then (2.38) reduces to the set of two equations

$$\frac{d}{dt}\left[\dot V_{(K')}S_Y^{\text{ext}}\right] + \left[\varsigma s_\tau(\ddot V_{(K')} + \Omega_0^2 V_{(K')}) - \varsigma s_n\dot V_{(K')}\dot\Phi\right]\sin\Phi +$$

$$+\left[\varsigma s_n(\ddot V_{(K')} + \Omega_0^2 V_{(K')}) + \varsigma s_\tau\dot V_{(K')}\dot\Phi\right]\cos\Phi + B_Y V_{(K')} = 0, \quad (2.45)$$

$$\frac{d}{dt}\left[\dot V_{(K')}S_X^{\text{ext}}\right] + \left[\varsigma s_\tau(\ddot V_{(K')} + \Omega_0^2 V_{(K')}) - \varsigma s_n\dot V_{(K')}\dot\Phi\right]\cos\Phi -$$

$$-\left[\varsigma s_n(\ddot V_{(K')} + \Omega_0^2 V_{(K')}) + \varsigma s_\tau\dot V_{(K')}\dot\Phi\right]\sin\Phi + B_X V_{(K')} = 0. \quad (2.46)$$

The energy (2.14) is conserved and may be represented as a sum of energies,

$$\mathcal{E} = \mathcal{E}^{(K')} + \mathcal{E}_0^{(K')} + \mathcal{E}_0, \quad (2.47)$$

where

$$\mathcal{E}^{(K')} = \frac{m_0 \mathbf{V}_{(K')}^2}{2} - \left(\mathbf{V}_{(K')} \cdot \frac{\partial U}{\partial \mathbf{V}_{(K')}}\right) + (\mathbf{V}_{(K')} \cdot [(\varsigma \mathbf{s} + \mathbf{S}^{\text{ext}}) \times \dot{\mathbf{V}}_{(K')}]) \tag{2.48}$$

is an energy of the r. f. $K'$,

$$\mathcal{E}_0 = \frac{m_0 \mathbf{v}^2}{2} - \left(\mathbf{v} \cdot \frac{\partial U}{\partial \mathbf{v}}\right) + U + (\mathbf{v} \cdot [(\varsigma \mathbf{s} + \mathbf{S}^{\text{ext}}) \times \dot{\mathbf{v}}]) =$$
$$= \frac{m_0 v^2}{2} - v\frac{du}{dv} + u - (\varsigma s_{\text{b}} + S_Z^{\text{ext}}) v^2 \dot{\Phi} \tag{2.49}$$

is the self-energy of the particle in the r. f. $K'$,

$$\mathcal{E}_0^{(K')} = m_0(\mathbf{v} \cdot \mathbf{V}_{(K')}) + (\mathbf{V}_{(K')} \cdot [(\varsigma \mathbf{s} + \mathbf{S}^{\text{ext}}) \times \dot{\mathbf{v}}]) + (\mathbf{v} \cdot [(\varsigma \mathbf{s} + \mathbf{S}^{\text{ext}}) \times \dot{\mathbf{V}}_{(K')}]) =$$
$$= V_{(K')} \left[ v\dot{\Phi}(\varsigma s_\tau + S_X^{\text{ext}} \cos\Phi + S_Y^{\text{ext}} \sin\Phi) - \dot{v}(\varsigma s_{\text{n}} - S_X^{\text{ext}} \sin\Phi + S_Y^{\text{ext}} \cos\Phi) \right] + \tag{2.50}$$
$$+ \dot{V}_{(K')} v(\varsigma s_{\text{n}} - S_X^{\text{ext}} \sin\Phi + S_Y^{\text{ext}} \cos\Phi)$$

is additional energy due to movement of the particle relative to r. f. $K'$.

From (2.49) it is easy to obtain the relation

$$(\varsigma s_{\text{b}} + S_Z^{\text{ext}})(v\ddot{\Phi} + 2\dot{v}\dot{\Phi}) + v\dot{S}_Z^{\text{ext}}\dot{\Phi} = \frac{d}{dt}\left(m_0 v - \frac{du}{dv}\right). \tag{2.51}$$

In view of (2.41)-(2.42), (2.51) and $\dot{V}_{(K')} = 0$ for massive particles (2.37) reduces to the set of equations

$$E_X \cos\Phi + E_Y \sin\Phi = 0, \tag{2.52}$$

$$\left(m_0 v - \frac{du}{dv}\right)\dot{\Phi} + (\varsigma s_{\text{b}} + S_Z^{\text{ext}})(\ddot{v} - v\dot{\Phi}^2) + \dot{S}_Z^{\text{ext}}\dot{v} + (\varsigma s_{\text{b}}\Omega_0^2 + B_Z)v = -E_X \sin\Phi + E_Y \cos\Phi, \tag{2.53}$$

$$\frac{d}{dt}\left[\varsigma s_\tau v\dot{\Phi} - \varsigma s_{\text{n}}\dot{v} + (S_X^{\text{ext}}\dot{v} + S_Y^{\text{ext}}v\dot{\Phi})\sin\Phi + (S_X^{\text{ext}}v\dot{\Phi} - S_Y^{\text{ext}}\dot{v})\cos\Phi\right] = E_Z. \tag{2.54}$$

For massless particles one should set $m_0 = 0$ in (2.53) and instead of (2.41)-(2.42) one should use (2.45)-(2.46). We emphasize that obtained system of equations holds for arbitrary external field produced by stationary sources, potential of interaction with which depends on the velocity of particle.

To conclude this section we note that the dynamic momentum (2.6) of the object can be represented as

$$\mathbf{P} = \mathbf{P}_{(K')} + \mathbf{p} + \mathbf{p}_s, \tag{2.55}$$

where

$$\mathbf{P}_{(K')} = m_0 \mathbf{V}_{(K')} + \varsigma[\mathbf{s} \times \dot{\mathbf{V}}_{(K')}] + [\mathbf{S}^{\text{ext}} \times \dot{\mathbf{V}}_{(K')}] \tag{2.56}$$

is a kinetic momentum of the r. f. $K'$ in absolute r. f. (with no interaction we have $\dot{\mathbf{V}}_{(K')} = \mathbf{0}$ and $\mathbf{P}_{(K')} = m_0 \mathbf{V}_{(K')}$),

$$\mathbf{p} = m_0 \mathbf{v} - \frac{\partial U}{\partial \mathbf{v}} + [\mathbf{S}^{\text{ext}} \times \dot{\mathbf{v}}] \tag{2.57}$$

is a dynamic momentum of the particle (excluding spin) in r. f. $K'$,

$$\mathbf{p}_s = \varsigma[\mathbf{s} \times \dot{\mathbf{v}}] \tag{2.58}$$

is a spin momentum that exists when the particle has spin.

# 3. The motion of free Particle

In [17]-[18] there considered equations of motion (2.1) and (2.12) for the case $U_0 = 0$, $\mathbf{S}^{\text{ext}} = \mathbf{0}$, $\mathbf{C}^{\text{ext}} = \mathbf{0}$, and all solutions of them are found. By definition an object will be free if the latter two conditions are fulfilled and the force (2.2) vanishes at all time. Then $U_0 = 0$ is special case of the condition $\partial U_0 / \partial \mathbf{R} = \mathbf{0}$, whence it follows that $U_0$ may be function of the velocity and accelerations. On the other hand, if free object will be defined by the condition $\partial U / \partial \mathbf{R} = \mathbf{0}$, whence $U = u(t, \mathbf{V}, \dot{\mathbf{V}}, \ddot{\mathbf{V}}, ..., \dot{\mathbf{V}}^{(N)})$, then in the l. h. s. of (2.1) a term (2.3) remains that makes sense of gyroscopic force. If $U$ does not depend explicitly on time, and velocity, and accelerations, then it is constant, and it may be set zero. It is easy to show in this case that joint solution of (2.1) and (2.12) leads to the fact that spin of the object is always collinear to the velocity $\mathbf{V}$ which remains constant, $\dot{\mathbf{V}} = \mathbf{0}$. Hence, the force (2.2) vanishes, and (2.5) gives $U = U_0$, so that the object in question moves inertially in total concordance with standard Newton's mechanics.

If $U = u \neq 0$, then one should be based on (2.52)-(2.54) and (2.41)-(2.42) for massive particles or (2.45)-(2.46) for massless particles at $\mathbf{E} = \mathbf{0}$, $\mathbf{B} = \mathbf{0}$ and $\mathbf{S}^{\text{ext}} = \mathbf{0}$. It follows from (2.41)-(2.42) that $s_\tau = 0$, $s_n = 0$, and (2.45)-(2.46) reduce to the set

$$s_\tau(\ddot{V}_{(K')} + \Omega_0^2 V_{(K')}) = s_n \dot{V}_{(K')} \dot{\Phi}, \tag{3.1}$$

$$s_n(\ddot{V}_{(K')} + \Omega_0^2 V_{(K')}) = -s_\tau \dot{V}_{(K')} \dot{\Phi}, \tag{3.2}$$

which are valid also when $s_\tau \neq 0$, $s_n \neq 0$. Equations (2.52)-(2.54) reduce to

$$\left(m_0 v - \frac{du}{dv}\right)\dot{\Phi} = -\varsigma s_b [\ddot{v} + (\Omega_0^2 - \dot{\Phi}^2) v], \tag{3.3}$$

$$\frac{d}{dt}\left[s_\tau v \dot{\Phi} - s_n \dot{v}\right] = 0, \tag{3.4}$$

where (3.4) takes place for massless particle and becomes identity for massive one.

Substitution of (2.51) at $S_Z^{\text{ext}} = 0$ into (3.3) leads to

$$\frac{d}{dt}\left(\frac{\ddot{v} + \Omega_0^2 v}{\dot{\Phi}}\right) + \dot{v}\dot{\Phi} = 0, \tag{3.5}$$

whence it follows the first integral

$$\frac{(\ddot{v} + \Omega_0^2 v)^2}{\dot{\Phi}^2} + \dot{v}^2 + \Omega_0^2 v^2 = D^2 = \text{const}, \tag{3.6}$$

and

$$\dot{\Phi} = \frac{\ddot{v} + \Omega_0^2 v}{\sqrt{D^2 - \dot{v}^2 - \Omega_0^2 v^2}}. \tag{3.7}$$

Now substituting (3.7) into (3.3), we obtain the equation

$$\frac{m_0 v - du/dv}{\sqrt{D^2 - \dot{v}^2 - \Omega_0^2 v^2}}(\ddot{v} + \Omega_0^2 v) = -\varsigma s_b (\ddot{v} + \Omega_0^2 v)\left[1 - \frac{(\ddot{v} + \Omega_0^2 v) v}{D^2 - \dot{v}^2 - \Omega_0^2 v^2}\right], \tag{3.8}$$

which is valid in two cases: I. $\ddot{v} + \Omega_0^2 v = 0$, $\dot{\Phi} = 0$, and II. $\ddot{v} + \Omega_0^2 v \neq 0$, $\dot{\Phi} \neq 0$. Let us consider them in details.

I.1. $m_0 \geq 0$, $\ddot{v} + \Omega_0^2 v = 0$, $\dot{\Phi} = 0$, $s_\tau = s_n = 0$. In this case we have

$$v = v_0 \cos(\Omega_0 t + \varphi_0), \tag{3.9}$$

whence it follows

$$\mathbf{v}(t) = v_0 \cos(\Omega_0 t + \varphi_0)(\cos\Phi \mathbf{e}_X + \sin\Phi \mathbf{e}_Y). \tag{3.10}$$

The equation of the trajectory in the absolute r. f. is given by

$$\mathbf{R}(t) = \mathbf{R}(0) + \frac{v_0}{\Omega_0}[\sin(\Omega_0 t + \varphi_0) - \sin\varphi_0](\cos\Phi \mathbf{e}_X + \sin\Phi \mathbf{e}_Y) + V_{(K')} t \mathbf{e}_Z, \tag{3.11}$$

i. e. the particle has a longitudinal spin polarization and oscillates in the plane (X,Y) around the center of inertia moving along the Z axis with velocity $\mathbf{V}_C = \mathbf{V}_{(K')}$. At $\Omega_0 = 0$ oscillations are absent, and particle moves along the Z axis in accordance with the law of inertia of Galileo-Newton.

I.2. $m_0 = 0$, $\ddot{v} + \Omega_0^2 v = 0$, $\dot{\Phi} = 0$. Equations (3.1) and (3.4) give $s_n = 0$, $s_\tau \neq 0$, $\ddot{V}_{(K')} + \Omega_0^2 V_{(K')} = 0$, $V_{(K')} = V_{(K')0} \cos(\Omega_0 t + \varphi_1)$. Thus, we obtain a trajectory

$$\mathbf{R}(t) = \mathbf{R}(0) + \frac{v_0}{\Omega_0}[\sin(\Omega_0 t + \varphi_0) - \sin\varphi_0](\cos\Phi \mathbf{e}_X + \sin\Phi \mathbf{e}_Y) + \\ + \frac{V_{(K')0}}{\Omega_0}[\sin(\Omega_0 t + \varphi_1) - \sin\varphi_1]\mathbf{e}_Z, \tag{3.12}$$

which in general is an ellipse, transforming into either circle at $\Delta\varphi = \varphi_1 - \varphi_0 = (2m+1)\pi/2$ or line segment at $\Delta\varphi = m\pi$, where $m$ is integer.

II. The second case of equation (3.8) for function $u(v)$ is specified by conditions $\ddot{v} + \Omega_0^2 v \neq 0$, $\dot{\Phi} \neq 0$, $s_\tau = s_n = 0$. We have

$$\frac{du}{dv} = \left(m_0 - \frac{\varsigma s_b(\ddot{v} + \Omega_0^2 v)}{\sqrt{D^2 - \dot{v}^2 - \Omega_0^2 v^2}}\right)v + \varsigma s_b\sqrt{D^2 - \dot{v}^2 - \Omega_0^2 v^2}. \tag{3.13}$$

Substituting (3.7) into (2.49), we obtain

$$\frac{d}{dv}\frac{u}{v} = \frac{1}{2}\left(m_0 - \frac{2\varsigma s_b(\ddot{v} + \Omega_0^2 v)}{\sqrt{D^2 - \dot{v}^2 - \Omega_0^2 v^2}}\right) - \frac{\mathcal{E}_0}{v^2}, \tag{3.14}$$

whence it follows

$$u(v) = \frac{m_0 v^2}{2} + \varsigma s_b v\sqrt{D^2 - \dot{v}^2 - \Omega_0^2 v^2} + \mathcal{E}_0. \tag{3.15}$$

Equation of motion has the infinite set of solutions, one of which corresponds to $\Phi = \Omega_D t$, where $\Omega_D = \pm\sqrt{\Omega_D^2} = \text{const}$. In this case equation (3.7) admits first integral of motion $\sqrt{D^2 - \dot{v}^2 - \Omega_0^2 v^2} + \Omega_D v = F = \text{const}$, which reduces to equation for velocity

$$\dot{v} = \pm\sqrt{D^2 - F^2 + 2F\Omega_D v - (\Omega_D^2 + \Omega_0^2)v^2}. \tag{3.16}$$

Substitution of (3.16) into (3.7) leads to equation

$$\ddot{v} + (\Omega_0^2 + \Omega_D^2)v = F\Omega_D, \tag{3.17}$$

which has general solution

$$v(t) = \frac{F\Omega_D}{\chi^2} + v_0\cos(\chi t + \varphi_0), \quad \chi = \sqrt{\Omega_D^2 + \Omega_0^2}, \tag{3.18}$$

$$\mathbf{v}(t) = \left(\frac{F\Omega_D}{\chi^2} + v_0\cos(\chi t + \varphi_0)\right)(\cos\Omega_D t \mathbf{e}_X + \sin\Omega_D t \mathbf{e}_Y). \tag{3.19}$$

It follows from (3.16) and (3.18), that velocity may vary in the limits $v_{min} \leq v \leq v_{max}$, where

$$v_{\min} = \frac{F\Omega_{\text{D}}}{\chi^2} - v_0 \geq 0, \; v_{\max} = \frac{F\Omega_{\text{D}}}{\chi^2} + v_0, \; v_0 = \frac{\sqrt{D^2\chi^2 - F^2\Omega_0^2}}{\chi^2}. \tag{3.20}$$

For $u(v)$ we obtain the expression

$$u(v) = \frac{m_0 v^2}{2} + \varsigma s_{\text{b}} v \,|\, F - \Omega_{\text{D}} v \,| + \mathcal{E}_0. \tag{3.21}$$

(2.18) and (3.19) lead to the law of motion for both massive and massless particles with longitudinal spin polarization

$$\mathbf{R}(t) = \mathbf{R}(0) + \frac{F}{\chi^2}\left[\sin\Omega_{\text{D}} t\, \mathbf{e}_X + (1 - \cos\Omega_{\text{D}} t)\mathbf{e}_Y\right] + V_{(\text{K}')} t\, \mathbf{e}_Z +$$

$$+ \frac{v_0}{2\Omega_0^2}\Big[(\chi - \Omega_{\text{D}})\sin[(\chi + \Omega_{\text{D}})t + \varphi_0] + (\chi + \Omega_{\text{D}})\sin[(\chi - \Omega_{\text{D}})t + \varphi_0] - 2\chi\sin\varphi_0\Big]\mathbf{e}_X -$$

$$- \frac{v_0}{2\Omega_0^2}\Big[(\chi - \Omega_{\text{D}})\cos[(\chi + \Omega_{\text{D}})t + \varphi_0] - (\chi + \Omega_{\text{D}})\cos[(\chi - \Omega_{\text{D}})t + \varphi_0] + 2\Omega_{\text{D}}\cos\varphi_0\Big]\mathbf{e}_Y.$$

(3.22)

Initial phase in (3.22) may be put zero, $\varphi_0 = 0$; then, $\varphi_0 \neq 0$ corresponds to rotation of the plot by the angle $\varphi_0$ in the plane (XY). In the center-of-inertia r. f. the motion is in a plane perpendicular to the direction of spin. In general, the trajectory is a rosette, the form of which is shown in Fig. 1 for $v_{\min} \neq 0$ and Fig. 2 for $v_{\min} = 0$.

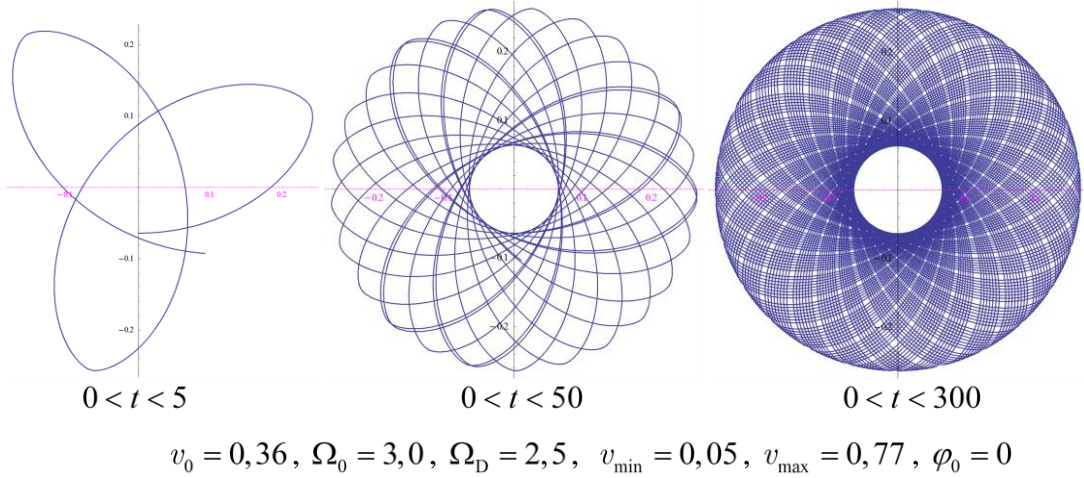

$0 < t < 5$     $0 < t < 50$     $0 < t < 300$

$v_0 = 0,36$, $\Omega_0 = 3,0$, $\Omega_{\text{D}} = 2,5$, $v_{\min} = 0,05$, $v_{\max} = 0,77$, $\varphi_0 = 0$

Figure 1. Types of trajectories (3.22) of free massive particle $v_{\min} < v < v_{\max}$

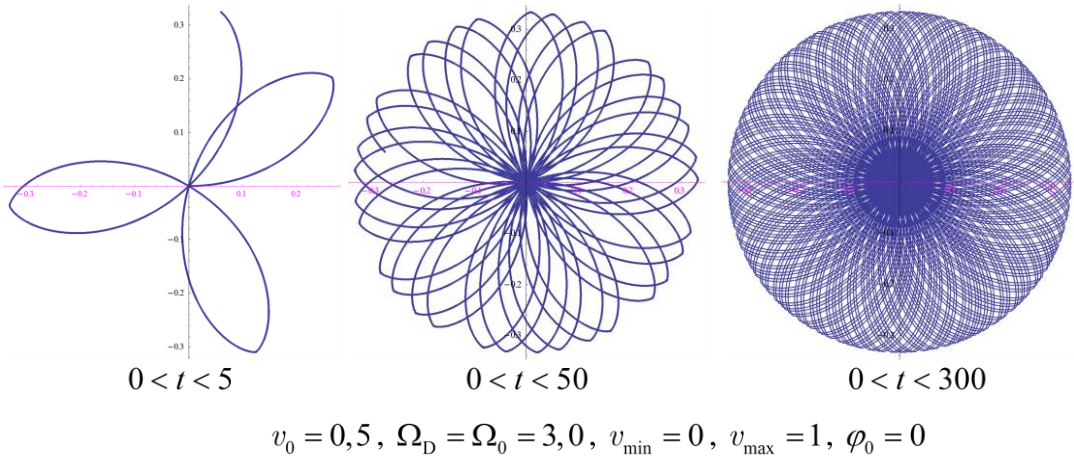

$0 < t < 5$     $0 < t < 50$     $0 < t < 300$

$v_0 = 0,5$, $\Omega_{\text{D}} = \Omega_0 = 3,0$, $v_{\min} = 0$, $v_{\max} = 1$, $\varphi_0 = 0$

Figure 2. Types of trajectories (3.22) of free massive particle $0 < v < v_{\max}$

The closed path can be obtained from the condition $\mathbf{v}(t+T) = \mathbf{v}(t)$, which leads to the condition $\Omega_D T = 2m\pi$, implying the ratio $m\chi = l\Omega_D$, or

$$m^2 \Omega_0^2 = (l^2 - m^2)\Omega_D^2, \tag{3.23}$$

where $l = 1, 2, \ldots$, $m = -l+1, -l+2, \ldots, 0, 1, 2, \ldots, l-1$, and $m = 0$ corresponds to $\Omega_D = 0$, i. e. to oscillations along the X axis. Examples of closed paths for $l = 2, 3, 4$ are shown in Fig. 3.

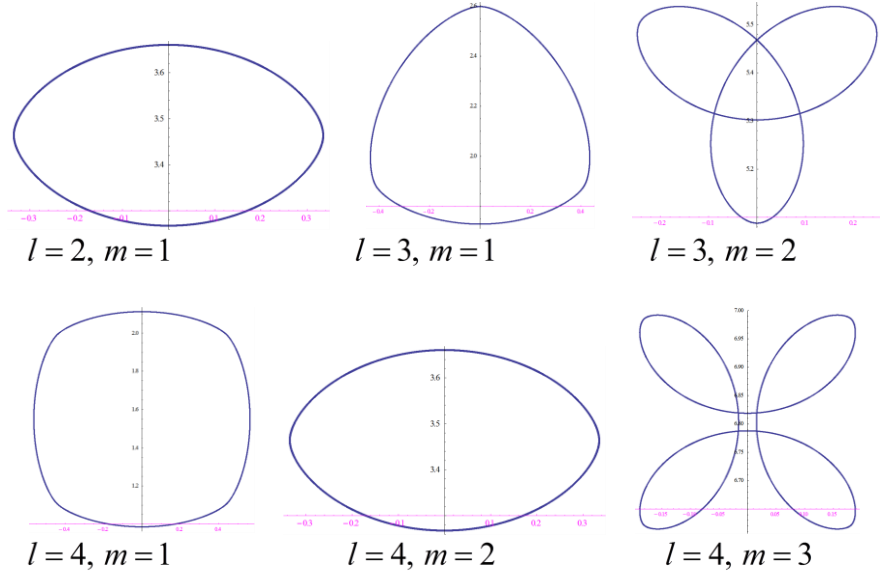

$l = 2, m = 1$      $l = 3, m = 1$      $l = 3, m = 2$

$l = 4, m = 1$      $l = 4, m = 2$      $l = 4, m = 3$

Figure 3. Shapes of closed trajectories (3.22) of free massive particle

Note that there is another possibility that the equation of motion (3.3)-(3.4) are satisfied.

III. $\dot{v} = 0$, $v = v_0 = \text{const}$. Then it follows from (2.49), (2.51)

$$\frac{d}{dt}\left(\frac{du}{dv} + \varsigma s_b v \dot{\Phi}\right) = 0, \quad \frac{du}{dv} + \varsigma s_b v \dot{\Phi} = C, \tag{3.24}$$

$$u(v) = \mathcal{E}_0 - \frac{m_0 v^2}{2} + v\left(\frac{du}{dv} + \varsigma s_b v \dot{\Phi}\right) = \mathcal{E}_0 - \frac{m_0 v^2}{2} + Cv. \tag{3.25}$$

Equation (3.3) reduces to

$$m_0 v + \varsigma s_b \frac{\Omega_0^2 v}{\dot{\Phi}} = \frac{du}{dv} + \varsigma s_b v \dot{\Phi} = C, \tag{3.26}$$

whence

$$\dot{\Phi} = \Omega_D = \frac{\varsigma s_b \Omega_0^2 v_0}{C - m_0 v_0} = \text{const}. \tag{3.27}$$

As a result we obtain following solutions.

III.1. $m_0 \geq 0$, $v = v_0$, $\Phi = \Omega_D t$, $s_\tau = s_n = 0$. Trajectory looks like helix along the Z axis

$$\mathbf{R}(t) = \mathbf{R}(0) + \frac{v_0}{\Omega_D}[\sin\Omega_D t \mathbf{e}_X + (1 - \cos\Omega_D t)\mathbf{e}_Y] + V_{(K')} t \mathbf{e}_Z. \tag{3.28}$$

De facto this solution coincides with solutions I.1, I.2 from [17], [18].

III.2. $m_0 = 0$, $v = v_0$, $\Phi = \Omega_D t$, $s_n = 0$. From (3.1) we find $\ddot{V}_{(K')} + \Omega_0^2 V_{(K')} = 0$, $V_{(K')} = V_{(K')0} \cos(\Omega_0 t + \varphi_1)$. Then the conservation of total energy (2.47) leads to $s_\tau = 0$. Trajectory is represented by radius vector

$$\mathbf{R}(t) = \mathbf{R}(0) + \frac{v_0}{\Omega_\mathrm{D}}[\sin\Omega_\mathrm{D}t\,\mathbf{e}_X + (1-\cos\Omega_\mathrm{D}t)\mathbf{e}_Y] + \frac{V_{(K')0}}{\Omega_0}[\sin(\Omega_0 t + \varphi_1) - \sin\varphi_1]\mathbf{e}_Z, \quad (3.29)$$

i. e. the particle performs complex movement around the stationary center of balance. At $\Omega_\mathrm{D} = \Omega_0$ the trajectory, which looks like three-dimensional Lissajous figure, becomes an ellipse (at $V_{(K')0} \neq v_0$) or circle (at $V_{(K')0} = v_0$ or $V_{(K')0} = 0$). The shape of non-elliptic trajectory is shown in Fig. 4.

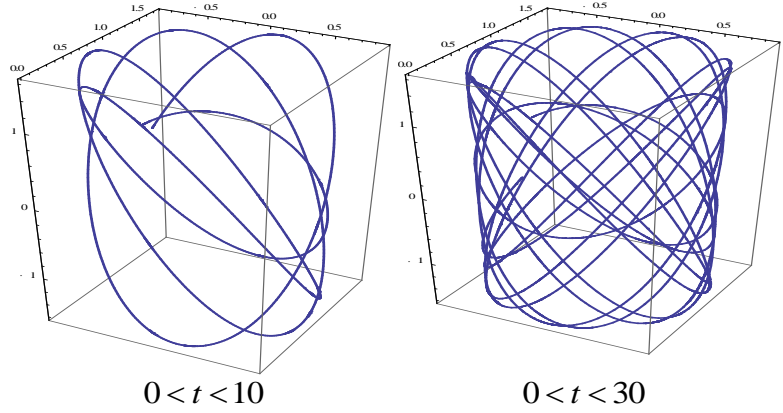

$0 < t < 10$   $0 < t < 30$

Figure 4. Type of trajectory (3.29) of free massless particle
$\Omega_\mathrm{D}/\Omega_0 = 1,2$, $V_{(K')0}/v_0 = 2,15$

### 4. Conclusion

We summarize the results of this section. We assume that in a homogeneous isotropic space behavior of a free spinning particle is described by equations (2.11) (with $\mathbf{E} = \mathbf{0}$, $\mathbf{B} = \mathbf{0}$) or (2.23) (with $\mathbf{E} = \mathbf{0}$, $\mathbf{B} = \mathbf{0}$, $\mathbf{S}^{\mathrm{ext}} = \mathbf{0}$) and (2.12). The problem of the dependence of potential function (2.15) and its physical sense requires special scrutiny. In this paper, solutions of equation of motion are found at quite natural simplifying assumption about the dependence of potential function on the velocity of the particle relative to an arbitrary moving center of inertia. As a result we have the following types of motion.

1. Oscillatory motion of massive or massless particle with longitudinal spin polarization in the plane, which is orthogonal to direction of movement of the center of inertia, described by radius vector (3.11). At $\Omega_0 = 0$ oscillations are absent and particle moves uniformly in a straight line. The self-energy $\mathcal{E}_0$ of the particle may have any constant value, $u(v) = m_0 v^2/2 + Cv + \mathcal{E}_0$.

2. Oscillatory or cyclic motion of massless particles around fixed center of balance with proper frequency $\Omega_0$, described by the law (3.12). Spin $\mathbf{s} = s_\tau(\cos\Phi\,\mathbf{e}_X + \sin\Phi\,\mathbf{e}_Y) + s_\mathrm{b}\mathbf{e}_Z$ has a constant direction. The self-energy $\mathcal{E}_0$ of the particle may have any constant value, $u(v) = Cv + \mathcal{E}_0$.

3. A motion of massive or massless particle with longitudinal spin polarization along complicated trajectory (3.22) clockwise ($\Omega_\mathrm{D} > 0$) or counter-clockwise ($\Omega_\mathrm{D} < 0$). The velocity of particle relative to the center of inertia varies in the limits $v_{\min} < v < v_{\max}$. Condition of closed trajectories leads to the relation (3.23), which implies that the integers *l* and *m* are as similar to the values of orbital and magnetic quantum numbers.

4. A motion of massive or massless particle with longitudinal spin polarization along helix (3.28) clockwise ($\Omega_\mathrm{D} > 0$, $s_\mathrm{b} < 0$) or counter-clockwise ($\Omega_\mathrm{D} < 0$, $s_\mathrm{b} > 0$) with constant ve-locity $v_0$ relative to the center of inertia, which moves in absolute r. f. with constant velocity $\mathbf{V}_C = V_{(K')}\mathbf{e}_Z$.

5. A motion of massless particle along complicated trajectory (3.29) around fixed center of balance. Spin $\mathbf{s} = s_b\mathbf{e}_Z$ is precessing around Z axis with angular velocity $\Omega_D$.

6. Besides the above types of motion, presumably, there may be other solutions corresponding to $\dot{\Phi} \neq \text{const}$. Potential function and function $\dot{\Phi}$ in a one-to-one manner determine each other. Their choice should be determined on the assumption that time interval of motion between two points of real path should be minimized, as required by Fermat's principle. However, it is not clear how it can be applied in this case.

In the cases above the movement comes from the infinite past to the infinite future. Most of the real problems is that a free particle flies in a certain region of space where the field is present, and then departs from it, again becomes free. Therefore, solutions of such problems can be obtained by combining the results of this paper with solutions that will be obtained in subsequent papers devoted to the motion of spinning particles in external fields.

Pay attention to the fact that the equation of motion allows the solution (3.12) and (3.29), corresponding to oscillations or movements of massless particles about fixed points balance . This suggests that the space can be filled with such particles, forming immovable Cartesian ether filled with vortices. However, discussions on this topic are beyond the scope of this article.

**Appendix A. The moving frame**

As it is known from differential geometry, with any trajectory of material point, moving in three-dimensional space, one can associate the moving frame, formed by three orthogonal unit vectors, called the tangent, principal normal and binormal. They are expressed as follows

$$\mathbf{e}_\tau = \frac{\mathbf{V}}{V} = \mathbf{R}' = [\mathbf{e}_n \times \mathbf{e}_b], \tag{A.1}$$

$$\mathbf{e}_n = \frac{[\mathbf{V} \times [\dot{\mathbf{V}} \times \mathbf{V}]]}{|[\mathbf{V} \times [\dot{\mathbf{V}} \times \mathbf{V}]]|} = \frac{V^2\dot{\mathbf{V}} - (\mathbf{V} \cdot \dot{\mathbf{V}})\mathbf{V}}{V|[\mathbf{V} \times \dot{\mathbf{V}}]|} =$$
$$= \frac{[\mathbf{R}' \times [\mathbf{R}'' \times \mathbf{R}']]}{|[\mathbf{R}' \times [\mathbf{R}'' \times \mathbf{R}']]|} = \frac{\mathbf{R}'' - (\mathbf{R}' \cdot \mathbf{R}'')\mathbf{R}'}{|[\mathbf{R}' \times \mathbf{R}'']|} = [\mathbf{e}_b \times \mathbf{e}_\tau], \tag{A.2}$$

$$\mathbf{e}_b = \frac{[\mathbf{V} \times \dot{\mathbf{V}}]}{|[\mathbf{V} \times \dot{\mathbf{V}}]|} = \frac{[\mathbf{R}' \times \mathbf{R}'']}{|[\mathbf{R}' \times \mathbf{R}'']|} = [\mathbf{e}_\tau \times \mathbf{e}_n], \tag{A.3}$$

respectively, where the point means differentiation with respect to time $t$, whereas the prime means differentiation with respect to natural parameter $s$ (length of trajectory arc), defined by

$$V = \frac{ds}{dt}, \quad s = \int_0^t V(t)dt = \int_0^t |\mathbf{V}(t)|\,dt, \tag{A.4}$$

so that $d/dt = (\dot{\,}) = Vd/ds = V(\,)'$.

Basis vectors (A.1)-(A.3) satisfy the orthogonality conditions
$$(\mathbf{e}_\tau \cdot \mathbf{e}_n) = 0, \ (\mathbf{e}_n \cdot \mathbf{e}_b) = 0, \ (\mathbf{e}_b \cdot \mathbf{e}_\tau) = 0 \tag{A.5}$$
and the Frenet-Serret equations

$$\frac{d\mathbf{e}_\tau}{dt} = \dot{\mathbf{e}}_\tau = V\mathbf{e}_\tau' = V\mathbf{R}'' = [\mathbf{\Omega}^D \times \mathbf{e}_\tau] = VK\mathbf{e}_n, \tag{A.6}$$

$$\frac{d\mathbf{e}_n}{dt} = \dot{\mathbf{e}}_n = V\mathbf{e}_n' = [\mathbf{\Omega}^D \times \mathbf{e}_n] = -VK\mathbf{e}_\tau + VT\mathbf{e}_b, \tag{A.7}$$

$$\frac{d\mathbf{e}_b}{dt} = \dot{\mathbf{e}}_b = V\mathbf{e}_b' = [\mathbf{\Omega}^D \times \mathbf{e}_b] = -VT\mathbf{e}_n, \tag{A.8}$$

where

$$\mathbf{\Omega}^{\mathrm{D}} = V(T\mathbf{e}_{\tau} + K\mathbf{e}_{\mathrm{b}}) = \frac{(\mathbf{V} \cdot [\dot{\mathbf{V}} \times \ddot{\mathbf{V}}])}{[\mathbf{V} \times \dot{\mathbf{V}}]^2} \mathbf{V} + \frac{[\mathbf{V} \times \dot{\mathbf{V}}]}{V^2} = V\left(\frac{(\mathbf{R}' \cdot [\mathbf{R}'' \times \mathbf{R}'''])}{[\mathbf{R}' \times \mathbf{R}'']^2} \mathbf{R}' + [\mathbf{R}' \times \mathbf{R}'']\right) \quad (A.9)$$

is Darboux vector defining the angular velocity of the moving frame,

$$K = \frac{1}{R_{\mathrm{K}}} = \frac{|[\mathbf{V} \times \dot{\mathbf{V}}]|}{V^3} = |[\mathbf{R}' \times \mathbf{R}'']| \quad (A.10)$$

is curvature, $R_{\mathrm{K}}$ is curvature radius of the trajectory in the given point,

$$T = \frac{1}{R_{\mathrm{T}}} = \frac{(\mathbf{V} \cdot [\dot{\mathbf{V}} \times \ddot{\mathbf{V}}])}{[\mathbf{V} \times \dot{\mathbf{V}}]^2} = \frac{(\mathbf{V} \cdot [\dot{\mathbf{V}} \times \ddot{\mathbf{V}}])}{V^6 K^2} = \frac{(\mathbf{R}' \cdot [\mathbf{R}'' \times \mathbf{R}'''])}{[\mathbf{R}' \times \mathbf{R}'']^2} = \frac{(\mathbf{R}' \cdot [\mathbf{R}'' \times \mathbf{R}'''])}{K^2} \quad (A.11)$$

is torsion, $R_{\mathrm{T}}$ is torsion radius of the trajectory in the given point,

$$K_f = \frac{|\mathbf{\Omega}^{\mathrm{D}}|}{V} = \sqrt{K^2 + T^2} \quad (A.12)$$

is total curvature of the trajectory.

Differentiation of radius vector $\mathbf{R}$ with respect to natural parameter gives

$$\mathbf{R}' = \mathbf{e}_{\tau} = \frac{\mathbf{V}}{V}, \quad (A.13)$$

$$\mathbf{R}'' = K\mathbf{e}_{\mathrm{n}} = -\frac{[\mathbf{V} \times [\mathbf{V} \times \dot{\mathbf{V}}]]}{V^4} = \frac{V^2 \dot{\mathbf{V}} - (\mathbf{V} \cdot \dot{\mathbf{V}})\mathbf{V}}{V^4}, \quad (A.14)$$

$$\mathbf{R}''' = K'\mathbf{e}_{\mathrm{n}} + K\mathbf{e}_{\mathrm{n}}' = K'\mathbf{e}_{\mathrm{n}} - K^2 \mathbf{e}_{\tau} + KT\mathbf{e}_{\mathrm{b}} = \frac{K'}{K}\mathbf{R}'' - K^2 \mathbf{R}' + KT\mathbf{e}_{\mathrm{b}} =$$
$$= -\frac{[\dot{\mathbf{V}} \times [\mathbf{V} \times \dot{\mathbf{V}}]]}{V^5} - \frac{[\mathbf{V} \times [\mathbf{V} \times \ddot{\mathbf{V}}]]}{V^5} + \frac{4(\mathbf{V} \cdot \dot{\mathbf{V}})[\mathbf{V} \times [\mathbf{V} \times \dot{\mathbf{V}}]]}{V^7} = \quad (A.15)$$
$$= \frac{\ddot{\mathbf{V}}}{V^3} - \frac{3(\mathbf{V} \cdot \dot{\mathbf{V}})\dot{\mathbf{V}}}{V^5} - \frac{\dot{\mathbf{V}}^2 \mathbf{V}}{V^5} - \frac{(\mathbf{V} \cdot \ddot{\mathbf{V}})\mathbf{V}}{V^5} + \frac{4(\mathbf{V} \cdot \dot{\mathbf{V}})^2 \mathbf{V}}{V^7}.$$

$$\mathbf{R}^{(4)} = K''\mathbf{e}_{\mathrm{n}} + K'\mathbf{e}_{\mathrm{n}}' - 2KK'\mathbf{e}_{\tau} - K^2 \mathbf{e}_{\tau}' + K'T\mathbf{e}_{\mathrm{b}} + KT'\mathbf{e}_{\mathrm{b}} + KT\mathbf{e}_{\mathrm{b}}' =$$
$$= -3KK'\mathbf{e}_{\tau} + (K'' - K^3 - KT^2)\mathbf{e}_{\mathrm{n}} + (2K'T + KT')\mathbf{e}_{\mathrm{b}} =$$
$$= \frac{2K'T + KT'}{KT}\mathbf{R}''' - \left(K^2 + T^2 + \frac{K'T' - K''T}{KT} + \frac{2K'^2}{K^2}\right)\mathbf{R}'' + \quad (A.16)$$
$$+ \frac{K}{T}(KT' - K'T)\mathbf{R}',$$

where

$$K' = \frac{([\mathbf{R}' \times \mathbf{R}''] \cdot [\mathbf{R}' \times \mathbf{R}'''])}{|[\mathbf{R}' \times \mathbf{R}'']|} = \frac{\mathbf{R}'^2 (\mathbf{R}'' \cdot \mathbf{R}''') - (\mathbf{R}' \cdot \mathbf{R}'')(\mathbf{R}' \cdot \mathbf{R}''')}{|[\mathbf{R}' \times \mathbf{R}'']|}, \quad (A.17)$$

$$\dot{K} = VK' = \frac{([\mathbf{V} \times \dot{\mathbf{V}}] \cdot [\mathbf{V} \times \ddot{\mathbf{V}}])}{V^3 |[\mathbf{V} \times \dot{\mathbf{V}}]|} - \frac{3(\mathbf{V} \cdot \dot{\mathbf{V}})|[\mathbf{V} \times \dot{\mathbf{V}}]|}{V^5} =$$
$$= \frac{V^4 (\dot{\mathbf{V}} \cdot \ddot{\mathbf{V}}) - V^2 (\mathbf{V} \cdot \dot{\mathbf{V}})(\mathbf{V} \cdot \ddot{\mathbf{V}}) - 3(\mathbf{V} \cdot \dot{\mathbf{V}})[\mathbf{V} \times \dot{\mathbf{V}}]^2}{V^5 |[\mathbf{V} \times \dot{\mathbf{V}}]|}; \quad (A.18)$$

$$T' = \frac{(\mathbf{R}' \cdot [\mathbf{R}'' \times \mathbf{R}^{(4)}])}{K^2} - \frac{2TK'}{K}, \quad (A.19)$$

$$\dot{T} = VT' = \frac{(\mathbf{V} \cdot [\dot{\mathbf{V}} \times \dddot{\mathbf{V}}])}{[\mathbf{V} \times \dot{\mathbf{V}}]^2} - \frac{2(\mathbf{V} \cdot [\dot{\mathbf{V}} \times \ddot{\mathbf{V}}])([\mathbf{V} \times \dot{\mathbf{V}}] \cdot [\mathbf{V} \times \ddot{\mathbf{V}}])}{[\mathbf{V} \times \dot{\mathbf{V}}]^4}. \quad (A.20)$$

$$K'' = \frac{[\mathbf{R}' \times \mathbf{R}''']^2 + ([\mathbf{R}' \times \mathbf{R}''] \cdot [\mathbf{R}'' \times \mathbf{R}''']) + ([\mathbf{R}' \times \mathbf{R}''] \cdot [\mathbf{R}' \times \mathbf{R}^{(4)}])}{|[\mathbf{R}' \times \mathbf{R}'']|} - $$
$$- \frac{([\mathbf{R}' \times \mathbf{R}''] \cdot [\mathbf{R}' \times \mathbf{R}'''])^2}{|[\mathbf{R}' \times \mathbf{R}'']|^3}.$$
(A.21)

Equation (A.16) is ordinary differential equation of the fourth order for radius vector **R**, which implies that the derivatives of **R** above the third order are expressed linearly in terms of derivatives of the first, second and third order.

For flat trajectories torsion vanishes, $T = 0$. Then (A.16) reduces to differential equation of the third order

$$\mathbf{R}''' = K'\mathbf{e}_n + K\mathbf{e}'_n = K'\mathbf{e}_n - K^2\mathbf{e}_\tau = \frac{K'}{K}\mathbf{R}'' - K^2\mathbf{R}',$$
(A.22)

which is equivalent to the second-order equation for the velocity **V**

$$\ddot{\mathbf{V}} = \frac{V^2(\dot{\mathbf{V}} \cdot \ddot{\mathbf{V}}) - (\mathbf{V} \cdot \dot{\mathbf{V}})(\mathbf{V} \cdot \ddot{\mathbf{V}})}{[\mathbf{V} \times \dot{\mathbf{V}}]^2}\dot{\mathbf{V}} + \frac{\dot{\mathbf{V}}^2(\mathbf{V} \cdot \ddot{\mathbf{V}}) - (\mathbf{V} \cdot \dot{\mathbf{V}})(\dot{\mathbf{V}} \cdot \ddot{\mathbf{V}})}{[\mathbf{V} \times \dot{\mathbf{V}}]^2}\mathbf{V}.$$
(A.23)